\documentclass[manuscript]{aastex62}          

\def\be{\begin{eqnarray}}
\def\ee{\end{eqnarray}}

\def\lsim{\;\raise0.3ex\hbox{$<$\kern-0.75em\raise-1.1ex\hbox{$\sim$}}\;}
\def\gsim{\;\raise0.3ex\hbox{$>$\kern-0.75em\raise-1.1ex\hbox{$\sim$}}\;}

\def\cmc{\rm ~cm^{-3}}
\def\kms{\rm ~km~s^{-1}}

\def\cmc{\rm ~cm^{-3}}
\def\diff{\rm ~cm^2~s^{-1}}
\def \kms {\rm ~km~s^{-1}}

\def\ergs{\rm ~erg~s^{-1}}

\def\enf{\rm ~erg~cm^{-2}~s^{-1}}

\def\arcsec{\hbox{$^{\prime\prime}$}}

\def\ecsb{erg cm$^{-2}$ s$^{-1}$ arcsec$^{-2}$ }
\def\ecsb2{erg cm$^{-2}$ s$^{-1}$ arcsec$^{-2}$}

\def\farcd{\hbox{$.\!\!^{\circ}$}}

\def\arcsec{\hbox{$^{\prime\prime}$}}

\def \ams {{\sl AMS-02 }}
\def \amsf {{\sl AMS-02}}
\def \dampe {{\sl DAMPE }}
\def \dampef {{\sl DAMPE}}
\def \calet {{\sl CALET }}
\def \caletf {{\sl CALET}}

\def \j0437{PSR~J0437-4715}

\shorttitle{ Accelerated Leptons from PSR~J0437-4715 }
\shortauthors{ A.M.~Bykov et al. }

\begin{document}

\title{ GeV-TeV Cosmic Ray Leptons in the Solar System 
from the Bow Shock Wind Nebula \\
of the Nearest Millisecond Pulsar J0437-4715 }

\author[0000-0003-0037-2288]{A.M.\ Bykov}
\affiliation{Ioffe Institute, 26 Politehnicheskaya, St.\ Petersburg, 194021, Russia}
\author[0000-0001-8356-9654]{A.E.\ Petrov}
\affiliation{Ioffe Institute, 26 Politehnicheskaya, St.\ Petersburg, 194021, Russia}
\author[0000-0001-7681-4316]{A.M.\ Krassilchtchikov}
\affiliation{Ioffe Institute, 26 Politehnicheskaya, St.\ Petersburg, 194021, Russia}
\author[0000-0001-8063-0034]{K.P.\ Levenfish}
\affiliation{Ioffe Institute, 26 Politehnicheskaya, St.\ Petersburg, 194021, Russia}
\author[0000-0001-8806-0259]{S.M.\ Osipov}
\affiliation{Ioffe Institute, 26 Politehnicheskaya, St.\ Petersburg, 194021, Russia}
\author[0000-0002-7481-5259]{G.G.\ Pavlov}
\affiliation{Department of Astronomy and Astrophysics, Pennsylvania State University, PA 16802, USA}
\correspondingauthor{A.M.\ Bykov}
\email{byk@astro.ioffe.ru}


\begin{abstract}
We consider acceleration of leptons up to GeV-TeV energies in the bow shock 
wind nebula of \j0437 and their subsequent diffusion through the interstellar magnetic fields.  
The leptons accelerated at the pulsar wind termination surface are 
injected into re-acceleration in colliding shock flows. Modelled
spectra of synchrotron emission from the accelerated electrons and positrons 
are consistent with the far-ultraviolet and X-ray observations of the nebula
carried out with the {\sl Hubble Space Telescope} and {\sl Chandra
X-ray Observatory}. These observations are employed to constrain the absolute 
fluxes of relativistic leptons, which are escaping from the nebula and
eventually reaching the Solar System after energy-dependent diffusion
through the local interstellar medium accompanied by synchrotron and Compton
losses. It is shown that accelerated leptons from the nebula of \j0437 
can be responsible both for the enhancement of the positron fraction above 
a few GeV detected by {\sl PAMELA} and \ams\ spectrometers and for the TeV
range lepton fluxes observed with {\sl H.E.S.S.},  {\sl VERITAS}, {\sl Fermi}, 
\caletf, and \dampef.
\end{abstract}



\section{Introduction}
\label{intro}

Recent precise in-orbit measurements of cosmic ray (CR) spectra in the GeV -- TeV range
performed by {\sl PAMELA}, \amsf, {\sl H.E.S.S.}, {\sl VERITAS}, \dampef, and \calet have revealed a non-trivial
structure of spectra of accelerated positrons and electrons 
\citep[][]{HESS2008,HESS2009,pamela09,Fermi2009,Fermi2012,2014PhRvL.113l1102A,ams14,ams19,2017Natur.552...63D,calet18,VERITAS2018}.
In particular, \citet{ams19} have conducted precise flux measurements of CR positrons 
at energies up to 1~TeV and concluded that the CR positron flux in this energy regime 
can be represented as a sum of two components: one produced by inelastic collisions of 
CR nuclei with the interstellar gas and dominating at low energies, and the other originating
from a yet unknown source and dominating at high energies up to about 800~GeV.
The second component shows a complex spectral behavior with a significant excess over the low energy 
flux, which is prominent from about 25~GeV and then approximately follows a power law 
up to 250-300~GeV. Inelastic interactions of the energetic hadronic 
component of galactic CRs with the nuclei of the local interstellar medium (ISM)
produce positrons as well as antiprotons and other secondary nuclei
\citep[see, e.g.,][and references therein]{1998ApJ...493..694M,2012ApJ...752...68V}. 
However, it seems difficult to understand the origin of the growth of 
the positron fraction in the CR leptons above 10~GeV due to the CR nuclei
interactions \citep[see, e.g.,][]{ams19}. 

The two alternatives to the origin of excess positrons as secondary CRs are
annihilations or decays of dark matter particles \citep[e.g.,][]{1984PhRvL..53..624S,bertone05,bergstrom08} 
and the presence of local sources of accelerated electrons and positrons, which are believed
to come from energetic pulsars and supernova remnants
\citep[][]{Atoyan_ea95,hooper09, malyshev09, yuksel09, amatoblasi11, Kisaka2012,profumo12}.

Pair acceleration by pulsars (both in the magnetosphere and at the termination shock)
was discussed as a possible source of the
observed positron excess above a few GeV.
\citet[][]{malyshev09} noted that acceleration at the pulsar 
wind (PW) termination shock is required to produce the multi-GeV 
positrons consistent with {\sl PAMELA} data. 

\citet{2008AdSpR..42..497B} and \citet{2015ApJ...807..130V} showed that pair 
cascades from the magnetospheres of
millisecond pulsars without wind nebulae could only modestly contribute to
the CR lepton fluxes near the Earth at a few tens of GeV, and this component would
cut off at higher energies. However, they pointed out that strong intrabinary
shocks in redback and black widow type pulsars may allow them to contribute
to 10-40 TeV cosmic ray fluxes near the Earth.
As the propagation distance of the accelerated $e^{\pm}$ pairs would
decrease with energy, at higher energies their spectrum should become
bump-like \citep[e.g.,][]{cholis18}. In any case, once observed with
a signifcant confidence, the lepton spectra at TeV energies
favor a major contribution from only one
or few local sources; otherwise, the bump would be smoothed away.

Because of the young estimated pulsar age of about 11 kyr,
the nearby Vela pulsar wind nebulae could contribute to the
observed lepton fluxes only if the diffusion coefficient for the particle
energies of interest were about 10$^{30} \diff$. This value is somewhat large
for the energy range 10 - 100 GeV, both for transport inside a supernova remnant and
for the local ISM  thus making a substantial contribution
from this pulsar to the observed lepton fluxes at 10 - 100 GeV unlikely. 
Above 100 GeV some possible contribution from the Vela-X  PWN  
was discussed \citep[see, e.g.,][]{2015JHEAp...8...27D}. 
The nearby middle-aged pulsars Geminga and PSR~B0656+14 could be 
potential sources of accelerated leptons 
\citep[see, e.g.,][]{2018ApJ...863...30F,2018PhRvD..97l3008P,2019MNRAS.484.3491T}. 
The nearby middle-aged PSR~B1055-52 may have a weak X-ray pulsar 
wind nebula \citep[][]{2015ApJ...811...96P}. The source has a spindown power 
similar to that of \j0437 (which we will discuss in detail here), but it is apparently farther away 
\citep{mignani10}.

The High-Altitude Water Cherenkov Observatory (HAWC) reported a
detection of extended TeV gamma-ray emission from Geminga and PSR~B0656+14
\citep{2017Sci...358..911A}. Analysing the observed gamma-ray emission
profiles with a single zone diffusion model, the HAWC team concluded that
the CR diffusion in the source vicinities is too slow 
for these pulsars to be responsible for
the positron excess observed by {\sl PAMELA} and \amsf.  However,
two-zone models with slow diffusion in the inner zone of about 40~pc around
the nebulae and fast diffusion in the local ISM
\citep{2018ApJ...863...30F,2018PhRvD..97l3008P,2019MNRAS.484.3491T} can explain,
under some conditions, both the TeV emission profiles and the observed
positron excess. The diffusion in the inner zone around a
pulsar wind nebula (the ``TeV halo'') can be strongly suppressed compared to the
background ISM due to non-linear effects of turbulence driven by CRs
escaping the source \citep[see, e.g.,][]{2013ApJ...768...73M,Evoli_2018}. This requires
a high pulsar spindown power and its efficient conversion to CR
pressure. To fit the \ams data, \citet{2018ApJ...863...30F} had to assume
at least 75\% efficiency of Geminga's spindown power conversion into 
energetic CRs. Based on a two zone model, \citet{2019MNRAS.484.3491T}
concluded that the Geminga pulsar wind nebula (PWN) could significantly
contribute to the observed positron excess above 300 GeV, while another
source is needed to provide the positrons between 10 and 300 GeV.
\citet{2018PhRvD..98d3005H} performed a stacked analysis of 24 old recycled
millisecond pulsars within the field of-view of the HAWC observatory. They
found evidence of the presence of TeV halos around these millisecond pulsars 
on the 2.6-3.1 $\sigma$ level.

An analysis presented by \citet[][]{2018PhRvL.121y1106L} suggested that an
undiscovered pulsar with the spindown power $\sim$10$^{33-34}~\ergs$ 
should exist in the 80~pc vicinity of the Solar System to explain the
observed TeV spectral feature, assuming a very low local CR
diffusion coefficient $\sim$ 10$^{26}\times(E/{\rm 10\ GeV})^{\rm
0.33}$~cm$^2$s$^{-1}$. On the other hand, \citet{2018arXiv181107551R}
pointed out that since the measured growth of the positron fraction above a few
GeV saturates at a level well below 0.5 and may even drop down above $\sim~400-500$~GeV 
(though the \ams data have rather large statistical errors at
these energies), the local sources of TeV leptons should not produce
positrons and electrons in equal amounts. \citet{2018arXiv181107551R}
claimed that this would rule out PWs as main sources of TeV range
leptons, and proposed instead the presence of a single local fading source of
TeV range electrons, which might be an old supernova remnant or another
object. Hence, to constrain the source models, one would need to carry out
high-precision measurements of the positron fraction at 0.5-2 TeV energies  
where the flux suppression (a spectral break feature) was revealed by \dampe and \caletf. 

When energetic pulsars powering PWNe move through
the ISM with supersonic velocities, they form bow shocks, which are often 
discovered via their H$\alpha$ emission \citep[see, e.g.,][and references therein]{br14}.  
While PWNe and bow shocks are rare among the old rotation-powered millisecond pulsars
\citep[e.g.,][]{bogdanov17},  they have been studied in some detail for a dozen
of young and middle-aged pulsars \citep[][]{karg17}. Such bow shock pulsar
wind nebulae (BSPWNe) are considered among main possible contributors to
the positron population of the Galaxy \citep{amatoblasi11,ssr17}.
Contrary to the slowly moving PWNe without bow shocks, BSPWNe can be the sites
of a specific efficient acceleration of particles in the colliding shock flow (CSF)
zone between the PW termination shock [or more generally,
termination surface (TS), see \citet{arons2012}] and the bow shock \citep{ssr17}.
Such acceleration is likely the cause for the hard spectra of synchrotron X-rays
observed in a number of BSPWNe [e.g., Geminga, see \citet{2017ApJ...835...66P}]. 
The CR lepton spectrum in such BSPWNe is likely to consist of two components,
one of them is due to $e^{\pm}$ pairs accelerated at the TS, and the other is formed
in the CSF region.

The particle spectrum $f\left(E\right) \propto E^{-s},~ s\sim 2.1 -
2.3$, formed in the vicinity of the PWN TS can naturally account for the
X-ray spectra of the Crab nebula \citep[see, e.g.,][]{arons2012} which has no
bow shock, as well as some other Crab-like objects. However, it cannot
explain the hard photon spectra of synchrotron X-rays detected from BSPWNe,
namely, the hard X-ray photon indices $\Gamma \sim 1$ of the lateral tails 
of the Geminga PWN \citep{2017ApJ...835...66P} and of the inner region of 
the Vela PWN \citep{KP08}. To explain these spectra, a second component 
with $s \sim 1$ at energies up to $E \sim$~100~TeV is required.  

\begin{figure}
\includegraphics[width=0.89\textwidth]{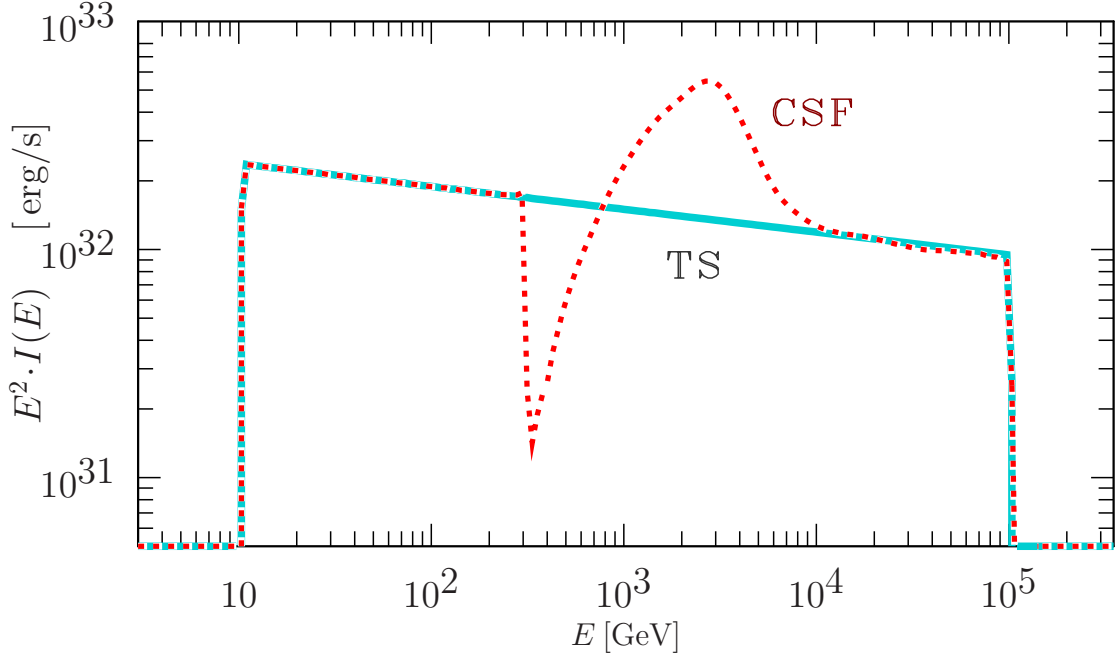}
\caption{A modelled spectrum of cosmic ray leptons escaping the nebula 
of \j0437 both through the tail and bow shock regions. The spectra formed at the termination surface (TS)
and in the colliding shock flow (CSF) are shown. The power
carried away by the CR particles leaving the modelled bow shock nebula
is about 30\% of the estimated $\dot{E}$. Here $I(E) = \int J(E) d\Omega$
is the direction-integrated spectral flux of the accelerated leptons.
}
\label{src_part}
\end{figure}

It was shown by \citet{ssr17} that starting at some high enough
energy, the $e^{\pm}$ pairs accelerated at the TS as well as
electrons and protons accelerated at the bow shock are injected into the
acceleration in the CSF, where they form 
a power law spectrum 
with $s<2$ in some energy range.
This can naturally explain the hard component of X-ray emission revealed in 
{\sl Chandra} observations of a number of BSPWNe.
It is important to note that the acceleration mechanism
operating in CSFs transfers the available wind flow energy 
upwards in the spectrum, i.e., to the $e^{\pm}$ pairs 
of the maximal energies that satisfy the condition of particle confinement in the
acceleration region. This substantially increases the efficiency of transfer of the
pulsar spindown power to the high energy pairs. 
In Figure~\ref{src_part} it is shown that the CRs accelerated in the 
CSFs provide the hard spectrum component between a few hundred GeV and a few TeV. 
The power law component at lower and higher energies 
is due to CR acceleration at the TS.  

Accelerated particles from nearby BSPWNe can reach the Solar System and are
likely to contribute to the locally measured spectra of galactic cosmic ray
(GCR) leptons. Below we discuss a model of particle acceleration in the
BSPWN of the nearest millisecond pulsar \j0437 and show that it can be 
the long-sought single source of the 10--800~GeV positron excess.
The CR leptons accelerated in the source can contribute also to the lepton spectrum at TeV 
energies measured in the Solar System, where the CR lepton flux suppression  (spectral break) 
was observed by both \calet and \dampe experiments \citep[][]{calet18}. 
The X-ray emission spectra observed from \j0437 nebula 
are used to calibrate the absolute lepton fluxes from this source.


\begin{figure}
\includegraphics[width=0.89\textwidth]{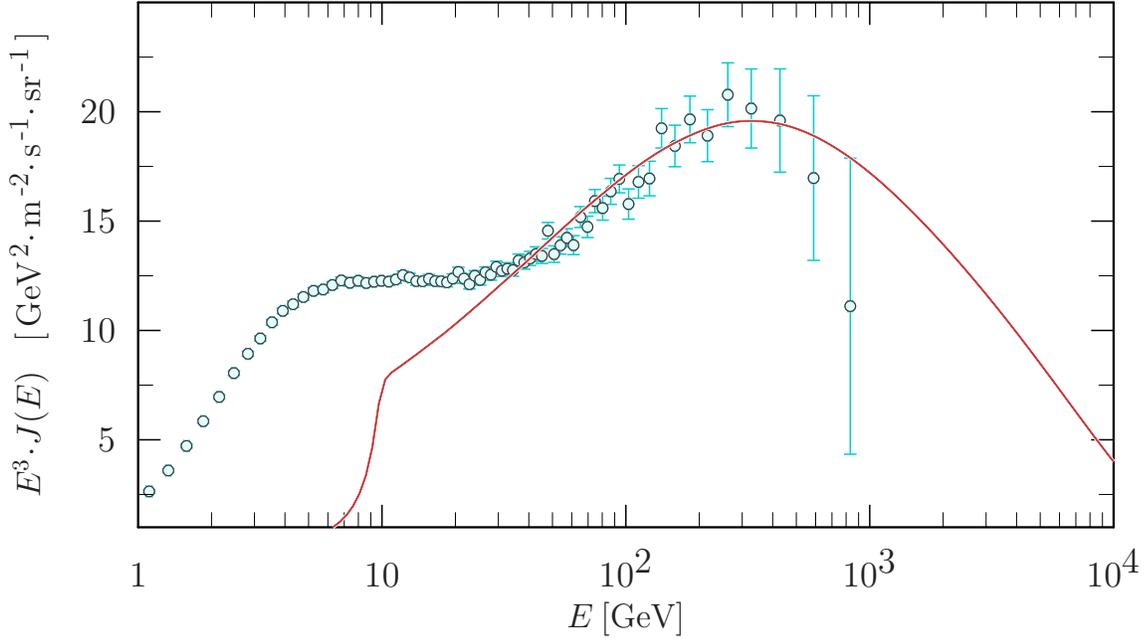}
\caption{A model spectrum of cosmic ray positrons (red) in the Solar System 
produced by the nebula of \j0437 after diffusion through the local interstellar
medium with account of synchrotron/Compton energy losses is shown
together with the observed spectrum \citep[blue;][]{ams19}. The considered 
diffusion is anisotropic, with the parallel and transverse diffusion coefficients  
$D_{\parallel} = 2 \times 10^{27} \left(E / 1 \mbox{~GeV}\right)^{1/3}$ cm$^2$\,s$^{-1}$ 
and $D_{\perp} = 0.03 D_{\parallel}$, respectively.
}
\label{ams_part}
\end{figure}


\begin{figure}
\includegraphics[width=0.89\textwidth]{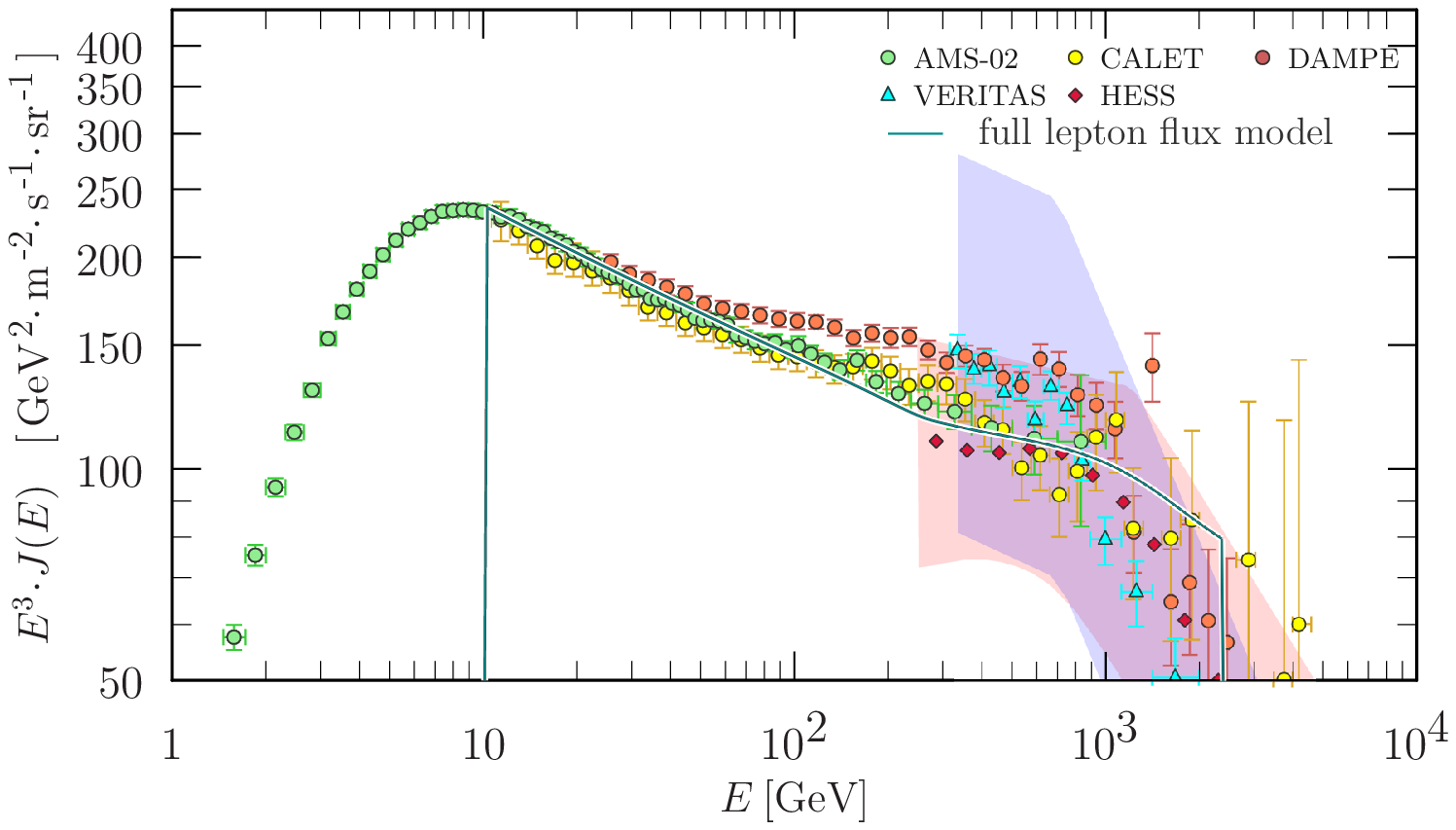}
\caption{The model spectrum of CR leptons originating from the nebula
of PSR~J0437-4715 confronted with the data measured 
by \amsf, {\sl H.E.S.S.},  \dampef, {\sl VERITAS}, and \calet  experiments (see Sect.~\ref{intro}). 
The pink and blue shaded regions indicate the systematic errors of {\sl H.E.S.S.} and {\sl VERITAS}, 
respectively.
}
\label{obs_mod}
\end{figure}


\section{Modeling of Particle Acceleration in the Nebula of PSR~J0437-4715}
\label{modeling}

The nebula of the old millisecond pulsar \j0437 was studied in the
optical, far-ultraviolet, and X-ray bands with the {\sl Hubble Space
Telescope} (HST) and {\sl Chandra X-ray Observatory} (CXO) by \citet{rangelov16}. 
\j0437 is located at a precisely measured distance of 156.79$\pm$0.25~pc \citep{reardon16}
in a binary system with a measured parallax. The transverse velocity of the system 
is $\approx$~104~$\kms$. 
The maximal realistic spindown power of the pulsar\footnote{
This value corresponds to a stiff equation of state, which allows the moment 
of inertia of a 1.44 $M_{\odot}$ neutron star to reach 2$\times$10$^{45}$ g$\cdot$cm$^2$.
}
corrected for the Shklovskii effect 
is $\dot{E}\sim$ 6$\times$10$^{33}~\ergs$, 
the number density of the ambient interstellar matter derived from observations
in the H$\alpha$ band is about 0.2~$\cmc$ \citep[see, e.g.,][]{br14}. 
The faint extended X-ray emission detected in the 5\arcsec\, vicinity of the
pulsar by \citet{rangelov16} is associated with the PWN, whose
X-ray luminosity $L_{X} \sim$ 3$\times$10$^{28}~\ergs$ and the 
photon index ${\rm \Gamma}=$ 1.8$\pm$0.4. 
The poorly constrained photon index allows 
both hard, $s \approx$1.8, and soft, $s \sim$3, indices of synchrotron radiating $e^{\pm}$ pairs
in the multi-TeV energy range. Far-ultraviolet (FUV) 
imaging revealed a bow shock structure with a 10\arcsec\, apex coinciding with 
the H$\alpha$ bow shock earlier observed by \citet{br14}. The unabsorbed 
1250 -- 2000 \AA\ luminosity of the bow shock $L_{\rm FUV} \sim 5\times 10^{28} \ergs$ 
is an order of magnitude higher than its H$\alpha$ luminosity 
\citep{rangelov16}. The observed FUV bow shock radiation can be produced by both
the heated interstellar gas emitting spectral lines such as C~IV 1549~\AA, 
O~IV 1403~\AA, Si~IV 1397~\AA, C~II 1335~\AA, He~II 1640~\AA\, and 
continuum synchrotron radiation of electrons and positrons of the PW.
To disentangle this contributions, FUV spectroscopy of the nebula is required.

Based on observations of young pulsars, such as the Crab pulsar, it is reasonable 
to assume that a nonthermal distribution 
of $e^{\pm}$ pairs of $f\left(E\right) \propto E^{-s},~ s\sim 2.1-2.3$,
which is responsible for the observed X-ray nebula, forms in the vicinity
of the wind TS of \j0437. An important issue is the composition of the 
PW: in addition to $e^{\pm}$ pairs, the wind may conatin a minor 
(by number density) but energetically significant ion component, which may
have important implications for modeling of the ultra high energy
CRs \citep[see, e.g.,][]{2015JCAP...07..016L,2015JCAP...08..026K}.
Particle-in-cell simulation of relativistic shocks in
electron-positron-proton plasmas with $m_p/m_e = 100$
\citep[][]{2006ApJ...653..325A}, demonstrated that the acceleration
efficiency and the spectra of the accelerated particles depend on the
plasma composition upstream of the shock. 1D simulations revealed
that if a sizeable fraction of the incoming energy flux is contained in protons,
the non-thermal distribution of $e^{\pm}$ pairs would form in the shock downstream 
with different spectral shapes of electrons and positrons. Namely, the positron spectra 
would be characterized by harder power law spectral indices, but lower cutoff energies
than those of the accelerated  electrons. In this case the positron fraction may vary 
with energy and be both above and below 0.5 in different energy ranges.   
For the case of \j0437 we assumed the dominance of electrons in the spectral cut off regime. 
As reliable models of PWs including the ion component are not available as of yet, 
we considered parameterized distributions of accelerated positrons and electrons at
the TS of the wind of \j0437. This parameterization assumed equal amounts
of electrons and positrons in the TS spectrum shown in Figure~\ref{src_part}, \
with an electron-dominated high energy end.

We used the Monte-Carlo modeling to simulate particle transport after
leaving the pulsar wind TS, where a power-law particle spectrum is thought
to be produced (see the curve labeled ``TS'' in Figure~\ref{src_part}). The
minimal Lorentz factor of the leptons accelerated at the TS was assumed to
be about the Lorentz factor of the cold pulsar wind of PSR J0437-4715. The
actual low energy cutoff might be rather smooth. However, we do not discuss
here the CR spectrum below 10 GeV affected by the solar wind modulation.
Thus we simply assume a sharp low-energy cutoff at $\sim$ 10 GeV. The
maximal energy is parameterized as a fraction of the total magnetospheric
potential $e \sqrt{\dot{E}/c}$  \citep[see, e.g.,][]{arons2012}. The energy
$cp_{\ast}$ at the peak of the lepton component re-accelerated at the CSF
(the red-dotted line labelled ``CSF'' in Figure \ref{src_part}) is
determined by the diffusion coefficient inside the bow shock PWN
$D(p_{\ast})$. This energy can be estimated from $D(p_{\ast}) \approx
u_{\rm psr} R_{\rm sh}$, where $u_{\rm psr}$ is the proper velocity of the
pulsar and $R_{\rm sh}$ is the size of the CSF accelerating region which is
about the apparent bow shock apex size \citep[for details, see][]{ssr17}.

In the Monte-Carlo simulation the injected TS spectrum
had a low-energy cutoff at the Lorenz factor $\gamma = \gamma_0 \sim (2-8) \times 10^5$. 
Particles with lower energies are not involved in the CSFs, they are
advected to the tail of the nebula by the PW flow. 

The escaping particle flux $J_{\rm calc}\left(E\right)$ and the energy
flux $\Phi_{\rm calc}\left(E\right)$ 
carried by the particles escaping the source were calculated within a Monte-Carlo
simulation (see Figure~\ref{src_part}). 
The flux of escaping particles was considered as the source term
in the transport equation, which described the CR lepton diffusion in the ISM
with synchrotron and Compton losses. This equation was solved to obtain 
the contribution of the BSPWN of \j0437 to the CR lepton flux observed near the Earth.

The Monte-Carlo technique allows one to calculate the particle distribution
function $f_{\rm calc}$ at a discrete spatial grid as well as the fluxes $J_{\rm calc}$,
$\Phi_{\rm calc}$ through a given surface surrounding the source.
These values are further rescaled to match the observed fluxes.
Namely, using the observed PWN X-ray flux, $F_{\rm obs} =
10^{-14}$ $\enf$, and the model flux value $F_{\rm calc} = \int I_{\nu}^{\rm
calc} \cos\theta d\Omega d\nu$, 
one can rescale the model particle flux $J$: 
\be
J = \frac{F_{\rm obs}}{F_{\rm calc}} J_{\rm calc}.
\ee
The value $F_{\rm calc}$ is obtained via a standard integration over
the solid angle $\Omega$ and frequency $\nu$ of $I_{\nu}^{\rm calc}$
-- the synchrotron emission intensity,
which is calculated along a given line of sight using the simulated local particle
spectra $f_{\rm calc}$ and standard formulae for the synchrotron emissivity
in the chaotic magnetic field \citep{1986A&A...164L..16C}.

To model the observed fluxes, at the first step the particle
distribution function is simulated in the entire volume of the modeled
source and the source synchrotron emission is calculated. 
After calculation the flux $F_{\rm calc}$ from the PWN, particle
and energy fluxes through the source boundary are calculated. Finally,
the transport equation with the source term determined by the spectrum 
of escaping CR leptons is solved. The model of particle transport from 
the source to the Solar System includes anisotropic CR diffusion in the 
ISM and CR particle energy losses due to synchrotron and inverse Compton 
radiation. The inverse Compton losses are accounted for with the 
approximation of \citet{2005MNRAS.363..954M}. 

CR transport from nearby sources located at the distance comparable with the 
coherence length of the interstellar magnetic field (about 100~pc) 
can differ from the global CR diffusion 
in the Galaxy \citep[see, e.g.,][]{2018MNRAS.473.4544S}.
In this study we employ anisotropic diffusion to consider the CR transport. 
This is because the CR particles at the energies of interest (below a few TeV) 
are highly magnetized, i.e. their gyroradii are much smaller than both 
their mean free paths and the coherence scale of the turbulent magnetic field $L_{\rm m}$. 
In this case the CR transport across the local ordered magnetic field can be described by the
transverse diffusion coefficient $D_{\perp}$, which is much smaller than the
parallel diffusion coefficient $D_{\parallel}$.

Detailed Monte-Carlo simulations of CR transport in chaotic
magnetic fields made by \citet{2002PhRvD..65b3002C} and \citet{2004JCAP...10..007C} show that 
in the case of Kolmogorov-type turbulence, which is supported by 
observations of interstellar magnetic fields 
at low rigidities $\rho = 2\pi r_g / L_{\rm m} < 1$, the diffusion coefficient $D_{\parallel} \propto
\rho^{1/3}$ while the ratio $D_{\perp} / D_{\parallel}$
does not depend on the CR particle energy. Here $r_g = E / e B$ is the particle gyroradius, $E$ and
$e$ are the particle energy and charge, $B$ is the mean magnetic field.
The condition $\rho < 1$ is well satisfied in the ISM with
typical magnetic field $B_{\rm ISM} \sim 3 ~\mu$G and $L_{\rm m} \sim 10^2$~pc. 
The energy dependence $D_{\parallel} \propto
\rho^{1/3}$ is  matching the inferred energy dependence of the global (average) CR
diffusion coefficient in the Galaxy, $D \approx 3 \times 10^{28}
\left(E / 1 \mbox{~GeV}\right)^{\delta}$~cm$^2$s$^{-1}$, $\delta
\sim 1/3$ \citep[see, e.g.,][]{2007ARNPS..57..285S}. 
Global diffusion averaged over the galactic scales 
can be described as nearly-isotropic diffusion, while the local diffusion 
at scales comparable to the coherence scale of the galactic magnetic 
field is highly anisotropic. Here we employed $D_{\perp}  = 0.03
D_{\parallel}$ in accordance with the results by \citet{2002PhRvD..65b3002C}.
 
The direction of the local ordered ISM magnetic field -- $l, b$ = 
36\farcd 2, 49\farcd 0
($\pm$16\farcd 0) -- was derived by \citet{2015ApJ...814..112F} from observations of
polarized starlight and from analysis of the {\sl IBEX} Ribbon observations. 
In the CR production and transport modeling presented in Figures~\ref{ams_part} and \ref{obs_mod} 
the local ordered magnetic field $B_{\rm ISM} = 3.7 \mu$G was taken to be directed
at $l =$52$^{\circ}$ $b= $ 49$^{\circ}$, which is consistent with \citet{2015ApJ...814..112F}
within the specified uncertainties.

The coefficient of parallel diffusion $D_{\parallel}(p)$ in the Local (super)Bubble 
is thought to be somewhat lower than the global average value refered above. 
Thus, in the anisotropic diffusion model, we adopt $D_{\parallel}(p) =$2$\times 10^{27} \left(E /
1 \mbox{~GeV}\right)^{1/3}$ cm$^2$\,s$^{-1}$, consistent with 
that suggested by \citet[][]{yuksel09,2018APh...102....1L,2019MNRAS.484.3491T}.
It should be noted that satisfactory fits to the observed fluxes shown
in Figures~\ref{ams_part} and~\ref{obs_mod} can be obtained for 
a wide parameter space of the ordered field direction and the value of diffusion
coefficient.
  
In Figure~\ref{obs_mod} a model spectrum of CR leptons 
in the Solar System is shown, with the contribution from \j0437 dominating
at high energies. The blue curve shows the sum of the modelled lepton flux 
from \j0437 with the contribution from other sources of CR leptons.  
The spectral shape of that contribution in the whole range from $\sim$10 GeV
to few TeV is given by power-law models with indices 2.94 and 3.25 for positrons
and electrons, respectively, matching the power-law fit of the low-energy
($\sim$ 15--30 GeV) component of the total CR lepton spectrum of
\citet{2014PhRvL.113l1102A}. 
The lepton flux measurements made with \amsf, {\sl H.E.S.S.}, {\sl
VERITAS}, \dampef, and \calet
\citep[see][]{HESS2008,ams19,2017Natur.552...63D,calet18,VERITAS2018} are
also shown in Figure \ref{obs_mod}.
The region colored in pink indicates the systematic errors of {\sl
H.E.S.S.} \footnote{see
https://www.mpi-hd.mpg.de/hfm/HESS/pages/home/som/2017/09/}, while the
region colored in blue is the same for {\sl VERITAS}. 

The positron fraction in the source spectrum
in Figure~\ref{src_part} was assumed to be 0.5 to simulate the fluxes measured
at the Earth's orbit, which are shown in Figure~\ref{ams_part}. 
If the positron fraction is different from this value in the source, the fluxes can be
scaled correspondingly. Moreover, the spectra of positrons and electrons at the TS
as well as their maximal momenta may differ if the ions are
energetically important in the pulsar wind according to the microscopic
simulations of \citet{2006ApJ...653..325A} discussed above. In this case the
CR lepton spectra in the flux suppression (spectral break)
regime in Figure~\ref{obs_mod} provided by \j0437 would be
dominated by TeV regime electrons consistent with the \ams positron data of 
\citet{ams19}. This important issue deserves further
investigation via 3D particle-in-cell modeling with realistic electron-to-ion
mass ratio. The flattening of the modelled total lepton spectrum above 200 GeV 
in Figure~\ref{obs_mod} is due to CR acceleration in the CSFs behind the bow 
shock of \j0437 (see the bump in Figure~\ref{src_part}).


\section{Summary}
\label{summ}

We have shown that the expected flux of the high energy $e^{\pm}$
pairs from the bow shock nebula of \j0437 can explain the enchanced $e^{+}/(e^{-} +
e^{+})$ ratio detected near the Earth with magnetic spectrometers {\sl PAMELA}
and \amsf, from about 10~GeV up to 800~GeV, as well as the flux suppression 
(spectral break) in the TeV range found with \calet and \dampef.

A distinctive feature of the suggested model is that the absolute fluxes of
the leptons accelerated in the nebula of \j0437 are derived from the model of its
synchrotron emission. The model employs the assumptions about
the structure of the plasma flows and the magnetic field strength in the nebula (of a few tens of $\mu$G) consistent with the recent numerical MHD simulations by
\citet{barkov19} and \citet{olmi19}. Comparison of the model predictions with the 
optical, far-ultraviolet, and X-ray observations of \j0437 and its nebula \citep[see][]{rangelov16}
allows us to estimate the absolute fluxes of CR leptons accelerated in the nebula.
The model is also capable of reproducing the far-ultraviolet and X-ray morphology
of the BSPWN. In particular, the synchrotron emission of the accelerated
$e^{\pm}$ pairs can explain the 1250--2000 \AA\
far-ultraviolet radiation of the bow shock region. This emphasizes the
importance of FUV spectroscopy of this nebula to separate the contribution
of the emission lines from the hot plasma in the bow shock downstream from
the synchrotron continuum. Within the suggested model, only about 30\% of the pulsar 
spindown power is required to be converted into accelerated $e^{\pm}$ pairs.

Although the model of CR lepton acceleration in the nebula of \j0437\ described above allows one to
reproduce the available observations of {\sl PAMELA}, \amsf, \dampef,
\caletf, {\sl H.E.S.S.} and {\sl VERITAS}, some contribution to the observed fluxes of accelerated leptons 
from other nearby pulsars (Geminga, PSR~B0656+14, PSR~B1055-52) can not be excluded.

\acknowledgements
Some of the modeling was performed at the ``Tornado'' subsystem of the supercomputing center
of St.~Petersburg Polytechnic University and at the JSCC RAS. 
A.M.B. and A.E.P. were supported by RSF grant 16-12-10225.

\bibliographystyle{apj}
\bibliographystyle{aasjournal}
\bibliography{j0437}

\begin{thebibliography}{}
\expandafter\ifx\csname natexlab\endcsname\relax\def\natexlab#1{#1}\fi
\providecommand{\url}[1]{\href{#1}{#1}}
\providecommand{\dodoi}[1]{doi:~\href{http://doi.org/#1}{\nolinkurl{#1}}}
\providecommand{\doeprint}[1]{\href{http://ascl.net/#1}{\nolinkurl{http://ascl.net/#1}}}
\providecommand{\doarXiv}[1]{\href{https://arxiv.org/abs/#1}{\nolinkurl{https://arxiv.org/abs/#1}}}

\bibitem[{{Abdo} {et~al.}(2009){Abdo}, {Ackermann}, {Ajello}, {Atwood},
  {Axelsson}, {Baldini}, {Ballet}, {Barbiellini}, {Bastieri}, {Battelino},
  {Baughman}, {Bechtol}, {Bellazzini}, {Berenji}, {Blandford}, {Bloom},
  {Bogaert}, {Bonamente}, {Borgland}, {Bregeon}, {Brez}, {Brigida}, {Bruel},
  {Burnett}, {Caliandro}, {Cameron}, {Caraveo}, {Carlson}, {Casandjian},
  {Cecchi}, {Charles}, {Chekhtman}, {Cheung}, {Chiang}, {Ciprini}, {Claus},
  {Cohen-Tanugi}, {Cominsky}, {Conrad}, {Cutini}, {Dermer}, {de Angelis}, {de
  Palma}, {Digel}, {di Bernardo}, {Do Couto E Silva}, {Drell}, {Dubois},
  {Dumora}, {Edmonds}, {Farnier}, {Favuzzi}, {Focke}, {Frailis}, {Fukazawa},
  {Funk}, {Fusco}, {Gaggero}, {Gargano}, {Gasparrini}, {Gehrels}, {Germani},
  {Giebels}, {Giglietto}, {Giordano}, {Glanzman}, {Godfrey}, {Grasso},
  {Grenier}, {Grondin}, {Grove}, {Guillemot}, {Guiriec}, {Hanabata}, {Harding},
  {Hartman}, {Hayashida}, {Hays}, {Hughes}, {J{\'o}hannesson}, {Johnson},
  {Johnson}, {Johnson}, {Kamae}, {Katagiri}, {Kataoka}, {Kawai}, {Kerr},
  {Kn{\"o}dlseder}, {Kocevski}, {Kuehn}, {Kuss}, {Lande}, {Latronico},
  {Lemoine-Goumard}, {Longo}, {Loparco}, {Lott}, {Lovellette}, {Lubrano},
  {Madejski}, {Makeev}, {Massai}, {Mazziotta}, {McConville}, {McEnery},
  {Meurer}, {Michelson}, {Mitthumsiri}, {Mizuno}, {Moiseev}, {Monte},
  {Monzani}, {Moretti}, {Morselli}, {Moskalenko}, {Murgia}, {Nolan}, {Norris},
  {Nuss}, {Ohsugi}, {Omodei}, {Orlando}, {Ormes}, {Ozaki}, {Paneque},
  {Panetta}, {Parent}, {Pelassa}, {Pepe}, {Pesce-Rollins}, {Piron}, {Pohl},
  {Porter}, {Profumo}, {Rain{\`o}}, {Rando}, {Razzano}, {Reimer}, {Reimer},
  {Reposeur}, {Ritz}, {Rochester}, {Rodriguez}, {Romani}, {Roth}, {Ryde},
  {Sadrozinski}, {Sanchez}, {Sander}, {Saz Parkinson}, {Scargle}, {Schalk},
  {Sellerholm}, {Sgr{\`o}}, {Smith}, {Smith}, {Spandre}, {Spinelli}, {Starck},
  {Stephens}, {Strickman}, {Strong}, {Suson}, {Tajima}, {Takahashi},
  {Takahashi}, {Tanaka}, {Thayer}, {Thayer}, {Thompson}, {Tibaldo}, {Tibolla},
  {Torres}, {Tosti}, {Tramacere}, {Uchiyama}, {Usher}, {van Etten},
  {Vasileiou}, {Vilchez}, {Vitale}, {Waite}, {Wallace}, {Wang}, {Winer},
  {Wood}, {Ylinen}, \& {Ziegler}}]{Fermi2009}
{Abdo}, A.~A., {Ackermann}, M., {Ajello}, M., {et~al.} 2009, Physical Review
  Letters, 102, 181101, \dodoi{10.1103/PhysRevLett.102.181101}

\bibitem[{{Abeysekara} {et~al.}(2017){Abeysekara}, {Albert}, {Alfaro},
  {Alvarez}, {{\'A}lvarez}, {Arceo}, {Arteaga-Vel{\'a}zquez}, {Avila Rojas},
  {Ayala Solares}, {Barber}, {Bautista-Elivar}, {Becerril}, {Belmont-Moreno},
  {BenZvi}, {Berley}, {Bernal}, {Braun}, {Brisbois}, {Caballero-Mora},
  {Capistr{\'a}n}, {Carrami{\~n}ana}, {Casanova}, {Castillo}, {Cotti},
  {Cotzomi}, {Couti{\~n}o de Le{\'o}n}, {De Le{\'o}n}, {De la Fuente},
  {Dingus}, {DuVernois}, {D{\'{\i}}az-V{\'e}lez}, {Ellsworth}, {Engel},
  {Enr{\'{\i}}quez-Rivera}, {Fiorino}, {Fraija}, {Garc{\'{\i}}a-Gonz{\'a}lez},
  {Garfias}, {Gerhardt}, {Gonz{\'a}lez Mu{\~n}oz}, {Gonz{\'a}lez}, {Goodman},
  {Hampel-Arias}, {Harding}, {Hern{\'a}ndez}, {Hern{\'a}ndez-Almada}, {Hinton},
  {Hona}, {Hui}, {H{\"u}ntemeyer}, {Iriarte}, {Jardin-Blicq}, {Joshi},
  {Kaufmann}, {Kieda}, {Lara}, {Lauer}, {Lee}, {Lennarz}, {Vargas},
  {Linnemann}, {Longinotti}, {Luis Raya}, {Luna-Garc{\'{\i}}a},
  {L{\'o}pez-Coto}, {Malone}, {Marinelli}, {Martinez}, {Martinez-Castellanos},
  {Mart{\'{\i}}nez-Castro}, {Mart{\'{\i}}nez-Huerta}, {Matthews},
  {Miranda-Romagnoli}, {Moreno}, {Mostaf{\'a}}, {Nellen}, {Newbold}, {Nisa},
  {Noriega-Papaqui}, {Pelayo}, {Pretz}, {P{\'e}rez-P{\'e}rez}, {Ren}, {Rho},
  {Rivi{\`e}re}, {Rosa-Gonz{\'a}lez}, {Rosenberg}, {Ruiz-Velasco}, {Salazar},
  {Salesa Greus}, {Sandoval}, {Schneider}, {Schoorlemmer}, {Sinnis}, {Smith},
  {Springer}, {Surajbali}, {Taboada}, {Tibolla}, {Tollefson}, {Torres},
  {Ukwatta}, {Vianello}, {Weisgarber}, {Westerhoff}, {Wisher}, {Wood},
  {Yapici}, {Yodh}, {Younk}, {Zepeda}, {Zhou}, {Guo}, {Hahn}, {Li}, \&
  {Zhang}}]{2017Sci...358..911A}
{Abeysekara}, A.~U., {Albert}, A., {Alfaro}, R., {et~al.} 2017, Science, 358,
  911, \dodoi{10.1126/science.aan4880}

\bibitem[{{Accardo} {et~al.}(2014){Accardo}, {Aguilar}, {Aisa}, {Alvino},
  {Ambrosi}, {Andeen}, {Arruda}, {Attig}, {Azzarello}, {Bachlechner}, \&
  et~al.}]{ams14}
{Accardo}, L., {Aguilar}, M., {Aisa}, D., {et~al.} 2014, Physical Review
  Letters, 113, 121101, \dodoi{10.1103/PhysRevLett.113.121101}

\bibitem[{{Ackermann} {et~al.}(2012){Ackermann}, {Ajello}, {Allafort},
  {Atwood}, {Baldini}, {Barbiellini}, {Bastieri}, {Bechtol}, {Bellazzini},
  {Berenji}, {Blandford}, {Bloom}, {Bonamente}, {Borgland}, {Bouvier},
  {Bregeon}, {Brigida}, {Bruel}, {Buehler}, {Buson}, {Caliandro}, {Cameron},
  {Caraveo}, {Casandjian}, {Cecchi}, {Charles}, {Chekhtman}, {Cheung},
  {Chiang}, {Ciprini}, {Claus}, {Cohen-Tanugi}, {Conrad}, {Cutini}, {de
  Angelis}, {de Palma}, {Dermer}, {Digel}, {Do Couto E Silva}, {Drell},
  {Drlica-Wagner}, {Favuzzi}, {Fegan}, {Ferrara}, {Focke}, {Fortin},
  {Fukazawa}, {Funk}, {Fusco}, {Gargano}, {Gasparrini}, {Germani}, {Giglietto},
  {Giommi}, {Giordano}, {Giroletti}, {Glanzman}, {Godfrey}, {Grenier}, {Grove},
  {Guiriec}, {Gustafsson}, {Hadasch}, {Harding}, {Hayashida}, {Hughes},
  {J{\'o}hannesson}, {Johnson}, {Kamae}, {Katagiri}, {Kataoka},
  {Kn{\"o}dlseder}, {Kuss}, {Lande}, {Latronico}, {Lemoine-Goumard}, {Llena
  Garde}, {Longo}, {Loparco}, {Lovellette}, {Lubrano}, {Madejski}, {Mazziotta},
  {McEnery}, {Michelson}, {Mitthumsiri}, {Mizuno}, {Moiseev}, {Monte},
  {Monzani}, {Morselli}, {Moskalenko}, {Murgia}, {Nakamori}, {Nolan}, {Norris},
  {Nuss}, {Ohno}, {Ohsugi}, {Okumura}, {Omodei}, {Orlando}, {Ormes}, {Ozaki},
  {Paneque}, {Parent}, {Pesce-Rollins}, {Pierbattista}, {Piron}, {Pivato},
  {Porter}, {Rain{\`o}}, {Rando}, {Razzano}, {Razzaque}, {Reimer}, {Reimer},
  {Reposeur}, {Ritz}, {Romani}, {Roth}, {Sadrozinski}, {Sbarra}, {Schalk},
  {Sgr{\`o}}, {Siskind}, {Spandre}, {Spinelli}, {Strong}, {Takahashi},
  {Takahashi}, {Tanaka}, {Thayer}, {Thayer}, {Tibaldo}, {Tinivella}, {Torres},
  {Tosti}, {Troja}, {Uchiyama}, {Usher}, {Vandenbroucke}, {Vasileiou},
  {Vianello}, {Vitale}, {Waite}, {Winer}, {Wood}, {Wood}, {Yang}, \&
  {Zimmer}}]{Fermi2012}
{Ackermann}, M., {Ajello}, M., {Allafort}, A., {et~al.} 2012, Physical Review
  Letters, 108, 011103, \dodoi{10.1103/PhysRevLett.108.011103}

\bibitem[{{Adriani} {et~al.}(2009){Adriani}, {Barbarino}, {Bazilevskaya},
  {Bellotti}, {Boezio}, {Bogomolov}, {Bonechi}, {Bongi}, {Bonvicini}, {Bottai},
  {Bruno}, {Cafagna}, {Campana}, {Carlson}, {Casolino}, {Castellini}, {de
  Pascale}, {de Rosa}, {de Simone}, {di Felice}, {Galper}, {Grishantseva},
  {Hofverberg}, {Koldashov}, {Krutkov}, {Kvashnin}, {Leonov}, {Malvezzi},
  {Marcelli}, {Menn}, {Mikhailov}, {Mocchiutti}, {Orsi}, {Osteria}, {Papini},
  {Pearce}, {Picozza}, {Ricci}, {Ricciarini}, {Simon}, {Sparvoli},
  {Spillantini}, {Stozhkov}, {Vacchi}, {Vannuccini}, {Vasilyev}, {Voronov},
  {Yurkin}, {Zampa}, {Zampa}, \& {Zverev}}]{pamela09}
{Adriani}, O., {Barbarino}, G.~C., {Bazilevskaya}, G.~A., {et~al.} 2009, \nat,
  458, 607, \dodoi{10.1038/nature07942}

\bibitem[{{Adriani} {et~al.}(2018){Adriani}, {Akaike}, {Asano}, {Asaoka},
  {Bagliesi}, {Berti}, {Bigongiari}, {Binns}, {Bonechi}, {Bongi}, {Brogi},
  {Buckley}, {Cannady}, {Castellini}, {Checchia}, {Cherry}, {Collazuol}, {di
  Felice}, {Ebisawa}, {Fuke}, {Guzik}, {Hams}, {Hareyama}, {Hasebe}, {Hibino},
  {Ichimura}, {Ioka}, {Ishizaki}, {Israel}, {Kasahara}, {Kataoka}, {Kataoka},
  {Katayose}, {Kato}, {Kawanaka}, {Kawakubo}, {Kohri}, {Krawczynski},
  {Krizmanic}, {Lomtadze}, {Maestro}, {Marrocchesi}, {Messineo}, {Mitchell},
  {Miyake}, {Moiseev}, {Mori}, {Mori}, {Mori}, {Motz}, {Munakata}, {Murakami},
  {Nakahira}, {Nishimura}, {de Nolfo}, {Okuno}, {Ormes}, {Ozawa}, {Pacini},
  {Palma}, {Papini}, {Penacchioni}, {Rauch}, {Ricciarini}, {Sakai}, {Sakamoto},
  {Sasaki}, {Shimizu}, {Shiomi}, {Sparvoli}, {Spillantini}, {Stolzi}, {Suh},
  {Sulaj}, {Takahashi}, {Takayanagi}, {Takita}, {Tamura}, {Tateyama},
  {Terasawa}, {Tomida}, {Torii}, {Tsunesada}, {Uchihori}, {Ueno}, {Vannuccini},
  {Wefel}, {Yamaoka}, {Yanagita}, {Yoshida}, {Yoshida}, \& {Calet
  Collaboration}}]{calet18}
{Adriani}, O., {Akaike}, Y., {Asano}, K., {et~al.} 2018, Physical Review
  Letters, 120, 261102, \dodoi{10.1103/PhysRevLett.120.261102}

\bibitem[{{Aguilar} {et~al.}(2014){Aguilar}, {Aisa}, {Alvino}, {Ambrosi},
  {Andeen}, {Arruda}, {Attig}, {Azzarello}, {Bachlechner}, {Barao}, \&
  et~al.}]{2014PhRvL.113l1102A}
{Aguilar}, M., {Aisa}, D., {Alvino}, A., {et~al.} 2014, Physical Review
  Letters, 113, 121102, \dodoi{10.1103/PhysRevLett.113.121102}

\bibitem[{{Aguilar} {et~al.}(2019){Aguilar}, {Ali Cavasonza}, {Ambrosi},
  {Arruda}, {Attig}, {Azzarello}, {Bachlechner}, {Barao}, {Barrau}, {Barrin},
  \& et~al.}]{ams19}
{Aguilar}, M., {Ali Cavasonza}, L., {Ambrosi}, G., {et~al.} 2019, Physical
  Review Letters, 122, 041102, \dodoi{10.1103/PhysRevLett.122.041102}

\bibitem[{{Aharonian} {et~al.}(2008){Aharonian}, {Akhperjanian}, {Barres de
  Almeida}, {Bazer-Bachi}, {Becherini}, {Behera}, {Benbow}, {Bernl{\"o}hr},
  {Boisson}, {Bochow}, {Borrel}, {Braun}, {Brion}, {Brucker}, {Brun},
  {B{\"u}hler}, {Bulik}, {B{\"u}sching}, {Boutelier}, {Carrigan}, {Chadwick},
  {Charbonnier}, {Chaves}, {Cheesebrough}, {Chounet}, {Clapson}, {Coignet},
  {Costamante}, {Dalton}, {Degrange}, {Deil}, {Dickinson}, {Djannati-Ata{\"i}},
  {Domainko}, {Drury}, {Dubois}, {Dubus}, {Dyks}, {Dyrda}, {Egberts},
  {Emmanoulopoulos}, {Espigat}, {Farnier}, {Feinstein}, {Fiasson},
  {F{\"o}rster}, {Fontaine}, {F{\"u}{\ss}ling}, {Gabici}, {Gallant},
  {G{\'e}rard}, {Giebels}, {Glicenstein}, {Gl{\"u}ck}, {Goret},
  {Hadjichristidis}, {Hauser}, {Hauser}, {Heinz}, {Heinzelmann}, {Henri},
  {Hermann}, {Hinton}, {Hoffmann}, {Hofmann}, {Holleran}, {Hoppe}, {Horns},
  {Jacholkowska}, {de Jager}, {Jung}, {Katarzy{\'n}ski}, {Kaufmann},
  {Kendziorra}, {Kerschhaggl}, {Khangulyan}, {Kh{\'e}lifi}, {Keogh}, {Komin},
  {Kosack}, {Lamanna}, {Lenain}, {Lohse}, {Marandon}, {Martin},
  {Martineau-Huynh}, {Marcowith}, {Maurin}, {McComb}, {Medina}, {Moderski},
  {Moulin}, {Naumann-Godo}, {de Naurois}, {Nedbal}, {Nekrassov}, {Niemiec},
  {Nolan}, {Ohm}, {Olive}, {de O{\~n}a Wilhelmi}, {Orford}, {Osborne},
  {Ostrowski}, {Panter}, {Pedaletti}, {Pelletier}, {Petrucci}, {Pita},
  {P{\"u}hlhofer}, {Punch}, {Quirrenbach}, {Raubenheimer}, {Raue}, {Rayner},
  {Renaud}, {Rieger}, {Ripken}, {Rob}, {Rosier-Lees}, {Rowell}, {Rudak},
  {Rulten}, {Ruppel}, {Sahakian}, {Santangelo}, {Schlickeiser}, {Sch{\"o}ck},
  {Schr{\"o}der}, {Schwanke}, {Schwarzburg}, {Schwemmer}, {Shalchi}, {Skilton},
  {Sol}, {Spangler}, {Stawarz}, {Steenkamp}, {Stegmann}, {Superina}, {Tam},
  {Tavernet}, {Terrier}, {Tibolla}, {van Eldik}, {Vasileiadis}, {Venter},
  {Vialle}, {Vincent}, {Vivier}, {V{\"o}lk}, {Volpe}, {Wagner}, {Ward},
  {Zdziarski}, \& {Zech}}]{HESS2008}
{Aharonian}, F., {Akhperjanian}, A.~G., {Barres de Almeida}, U., {et~al.} 2008,
  Physical Review Letters, 101, 261104, \dodoi{10.1103/PhysRevLett.101.261104}

\bibitem[{{Aharonian} {et~al.}(2009){Aharonian}, {Akhperjanian}, {Anton},
  {Barres de Almeida}, {Bazer-Bachi}, {Becherini}, {Behera}, {Bernl{\"o}hr},
  {Bochow}, {Boisson}, {Bolmont}, {Borrel}, {Brucker}, {Brun}, {Brun},
  {B{\"u}hler}, {Bulik}, {B{\"u}sching}, {Boutelier}, {Chadwick},
  {Charbonnier}, {Chaves}, {Cheesebrough}, {Chounet}, {Clapson}, {Coignet},
  {Dalton}, {Daniel}, {Davids}, {Degrange}, {Deil}, {Dickinson},
  {Djannati-Ata{\"i}}, {Domainko}, {O'C.~Drury}, {Dubois}, {Dubus}, {Dyks},
  {Dyrda}, {Egberts}, {Emmanoulopoulos}, {Espigat}, {Farnier}, {Feinstein},
  {Fiasson}, {F{\"o}rster}, {Fontaine}, {F{\"u}{\ss}ling}, {Gabici}, {Gallant},
  {G{\'e}rard}, {Gerbig}, {Giebels}, {Glicenstein}, {Gl{\"u}ck}, {Goret},
  {G{\"o}ring}, {Hauser}, {Hauser}, {Heinz}, {Heinzelmann}, {Henri}, {Hermann},
  {Hinton}, {Hoffmann}, {Hofmann}, {Holleran}, {Hoppe}, {Horns},
  {Jacholkowska}, {de Jager}, {Jahn}, {Jung}, {Katarzy{\'n}ski}, {Katz},
  {Kaufmann}, {Kendziorra}, {Kerschhaggl}, {Khangulyan}, {Kh{\'e}lifi},
  {Keogh}, {Klu{\'z}niak}, {Kneiske}, {Komin}, {Kosack}, {Kossakowski},
  {Lamanna}, {Lenain}, {Lohse}, {Marandon}, {Martin}, {Martineau-Huynh},
  {Marcowith}, {Masbou}, {Maurin}, {McComb}, {Medina}, {Moderski}, {Moulin},
  {Naumann-Godo}, {de Naurois}, {Nedbal}, {Nekrassov}, {Nicholas}, {Niemiec},
  {Nolan}, {Ohm}, {Olive}, {de O{\~n}a Wilhelmi}, {Orford}, {Ostrowski},
  {Panter}, {Paz Arribas}, {Pedaletti}, {Pelletier}, {Petrucci}, {Pita},
  {P{\"u}hlhofer}, {Punch}, {Quirrenbach}, {Raubenheimer}, {Raue}, {Rayner},
  {Reimer}, {Renaud}, {Rieger}, {Ripken}, {Rob}, {Rosier-Lees}, {Rowell},
  {Rudak}, {Rulten}, {Ruppel}, {Sahakian}, {Santangelo}, {Schlickeiser},
  {Sch{\"o}ck}, {Schr{\"o}der}, {Schwanke}, {Schwarzburg}, {Schwemmer},
  {Shalchi}, {Sikora}, {Skilton}, {Sol}, {Spangler}, {Stawarz}, {Steenkamp},
  {Stegmann}, {Stinzing}, {Superina}, {Szostek}, {Tam}, {Tavernet}, {Terrier},
  {Tibolla}, {Tluczykont}, {van Eldik}, {Vasileiadis}, {Venter}, {Venter},
  {Vialle}, {Vincent}, {Vivier}, {V{\"o}lk}, {Volpe}, {Wagner}, {Ward},
  {Zdziarski}, \& {Zech}}]{HESS2009}
{Aharonian}, F., {Akhperjanian}, A.~G., {Anton}, G., {et~al.} 2009, \aap, 508,
  561, \dodoi{10.1051/0004-6361/200913323}

\bibitem[{{Amato} \& {Arons}(2006)}]{2006ApJ...653..325A}
{Amato}, E., \& {Arons}, J. 2006, \apj, 653, 325, \dodoi{10.1086/508050}

\bibitem[{{Archer} {et~al.}(2018){Archer}, {Benbow}, {Bird}, {Brose},
  {Buchovecky}, {Buckley}, {Bugaev}, {Connolly}, {Cui}, {Daniel}, {Feng},
  {Finley}, {Fortson}, {Furniss}, {Gillanders}, {H{\"u}tten}, {Hanna},
  {Hervet}, {Holder}, {Hughes}, {Humensky}, {Johnson}, {Kaaret}, {Kar},
  {Kelley-Hoskins}, {Kertzman}, {Kieda}, {Krause}, {Krennrich}, {Kumar},
  {Lang}, {Lin}, {Maier}, {McArthur}, {Moriarty}, {Mukherjee}, {O'Brien},
  {Ong}, {Otte}, {Petrashyk}, {Pohl}, {Pueschel}, {Quinn}, {Ragan}, {Reynolds},
  {Richards}, {Roache}, {Rulten}, {Sadeh}, {Santander}, {Sembroski}, {Staszak},
  {Sushch}, {Wakely}, {Wells}, {Wilcox}, {Wilhelm}, {Williams}, {Williamson},
  {Zitzer}, \& {VERITAS Collaboration}}]{VERITAS2018}
{Archer}, A., {Benbow}, W., {Bird}, R., {et~al.} 2018, \prd, 98, 062004,
  \dodoi{10.1103/PhysRevD.98.062004}

\bibitem[{{Arons}(2012)}]{arons2012}
{Arons}, J. 2012, \ssr, 173, 341, \dodoi{10.1007/s11214-012-9885-1}

\bibitem[{{Atoyan} {et~al.}(1995){Atoyan}, {Aharonian}, \&
  {V{\"o}lk}}]{Atoyan_ea95}
{Atoyan}, A.~M., {Aharonian}, F.~A., \& {V{\"o}lk}, H.~J. 1995, \prd, 52, 3265,
  \dodoi{10.1103/PhysRevD.52.3265}

\bibitem[{{Barkov} {et~al.}(2019){Barkov}, {Lyutikov}, \&
  {Khangulyan}}]{barkov19}
{Barkov}, M.~V., {Lyutikov}, M., \& {Khangulyan}, D. 2019, \mnras, 484, 4760,
  \dodoi{10.1093/mnras/stz213}

\bibitem[{{Bergstr{\"o}m} {et~al.}(2008){Bergstr{\"o}m}, {Bringmann}, \&
  {Edsj{\"o}}}]{bergstrom08}
{Bergstr{\"o}m}, L., {Bringmann}, T., \& {Edsj{\"o}}, J. 2008, \prd, 78,
  103520, \dodoi{10.1103/PhysRevD.78.103520}

\bibitem[{{Bertone} {et~al.}(2005){Bertone}, {Hooper}, \& {Silk}}]{bertone05}
{Bertone}, G., {Hooper}, D., \& {Silk}, J. 2005, \physrep, 405, 279,
  \dodoi{10.1016/j.physrep.2004.08.031}

\bibitem[{{Blasi} \& {Amato}(2011)}]{amatoblasi11}
{Blasi}, P., \& {Amato}, E. 2011, Astrophysics and Space Science Proc., 21,
  624.
\newblock \doarXiv{1007.4745}

\bibitem[{{Bogdanov}(2017)}]{bogdanov17}
{Bogdanov}, S. 2017, in Astrophysics and Space Science Library, Vol. 446,
  Modelling Pulsar Wind Nebulae, ed. D.~F. {Torres}, 295

\bibitem[{{Brownsberger} \& {Romani}(2014)}]{br14}
{Brownsberger}, S., \& {Romani}, R.~W. 2014, \apj, 784, 154,
  \dodoi{10.1088/0004-637X/784/2/154}

\bibitem[{{B{\"u}sching} {et~al.}(2008){B{\"u}sching}, {Venter}, \& {de
  Jager}}]{2008AdSpR..42..497B}
{B{\"u}sching}, I., {Venter}, C., \& {de Jager}, O.~C. 2008, Advances in Space
  Research, 42, 497, \dodoi{10.1016/j.asr.2007.08.005}

\bibitem[{{Bykov} {et~al.}(2017){Bykov}, {Amato}, {Petrov}, {Krassilchtchikov},
  \& {Levenfish}}]{ssr17}
{Bykov}, A.~M., {Amato}, E., {Petrov}, A.~E., {Krassilchtchikov}, A.~M., \&
  {Levenfish}, K.~P. 2017, \ssr, 207, 235, \dodoi{10.1007/s11214-017-0371-7}

\bibitem[{{Candia} \& {Roulet}(2004)}]{2004JCAP...10..007C}
{Candia}, J., \& {Roulet}, E. 2004, \jcap, 10, 007,
  \dodoi{10.1088/1475-7516/2004/10/007}

\bibitem[{{Casse} {et~al.}(2002){Casse}, {Lemoine}, \&
  {Pelletier}}]{2002PhRvD..65b3002C}
{Casse}, F., {Lemoine}, M., \& {Pelletier}, G. 2002, \prd, 65, 023002,
  \dodoi{10.1103/PhysRevD.65.023002}

\bibitem[{{Cholis} {et~al.}(2018){Cholis}, {Karwal}, \&
  {Kamionkowski}}]{cholis18}
{Cholis}, I., {Karwal}, T., \& {Kamionkowski}, M. 2018, \prd, 97, 123011,
  \dodoi{10.1103/PhysRevD.97.123011}

\bibitem[{{Crusius} \& {Schlickeiser}(1986)}]{1986A&A...164L..16C}
{Crusius}, A., \& {Schlickeiser}, R. 1986, \aap, 164, L16

\bibitem[{{DAMPE Collaboration}(2017)}]{2017Natur.552...63D}
{DAMPE Collaboration}. 2017, \nat, 552, 63, \dodoi{10.1038/nature24475}

\bibitem[{{Della Torre} {et~al.}(2015){Della Torre}, {Gervasi}, {Rancoita},
  {Rozza}, \& {Treves}}]{2015JHEAp...8...27D}
{Della Torre}, S., {Gervasi}, M., {Rancoita}, P.~G., {Rozza}, D., \& {Treves},
  A. 2015, Journal of High Energy Astrophysics, 8, 27,
  \dodoi{10.1016/j.jheap.2015.08.001}

\bibitem[{{Evoli} {et~al.}(2018){Evoli}, {Linden}, \& {Morlino}}]{Evoli_2018}
{Evoli}, C., {Linden}, T., \& {Morlino}, G. 2018, \prd, 98, 063017,
  \dodoi{10.1103/PhysRevD.98.063017}

\bibitem[{{Fang} {et~al.}(2018){Fang}, {Bi}, {Yin}, \&
  {Yuan}}]{2018ApJ...863...30F}
{Fang}, K., {Bi}, X.-J., {Yin}, P.-F., \& {Yuan}, Q. 2018, \apj, 863, 30,
  \dodoi{10.3847/1538-4357/aad092}

\bibitem[{{Frisch} {et~al.}(2015){Frisch}, {Berdyugin}, {Piirola}, {Magalhaes},
  {Seriacopi}, {Wiktorowicz}, {Andersson}, {Funsten}, {McComas}, {Schwadron},
  {Slavin}, {Hanson}, \& {Fu}}]{2015ApJ...814..112F}
{Frisch}, P.~C., {Berdyugin}, A., {Piirola}, V., {et~al.} 2015, \apj, 814, 112,
  \dodoi{10.1088/0004-637X/814/2/112}

\bibitem[{{Hooper} {et~al.}(2009){Hooper}, {Blasi}, \& {Serpico}}]{hooper09}
{Hooper}, D., {Blasi}, P., \& {Serpico}, P.~D. 2009, \jcap, 1, 025,
  \dodoi{10.1088/1475-7516/2009/01/025}

\bibitem[{{Hooper} \& {Linden}(2018)}]{2018PhRvD..98d3005H}
{Hooper}, D., \& {Linden}, T. 2018, \prd, 98, 043005,
  \dodoi{10.1103/PhysRevD.98.043005}

\bibitem[{{Kargaltsev} \& {Pavlov}(2008)}]{KP08}
{Kargaltsev}, O., \& {Pavlov}, G.~G. 2008, in AIP Conf. Series, Vol. 983, 40
  Years of Pulsars: Millisecond Pulsars, Magnetars and More, ed. C.~{Bassa},
  Z.~{Wang}, A.~{Cumming}, \& V.~M. {Kaspi}, 171--185

\bibitem[{{Kargaltsev} {et~al.}(2017){Kargaltsev}, {Pavlov}, {Klingler}, \&
  {Rangelov}}]{karg17}
{Kargaltsev}, O., {Pavlov}, G.~G., {Klingler}, N., \& {Rangelov}, B. 2017,
  Journal of Plasma Physics, 83, 635830501, \dodoi{10.1017/S0022377817000630}

\bibitem[{{Kisaka} \& {Kawanaka}(2012)}]{Kisaka2012}
{Kisaka}, S., \& {Kawanaka}, N. 2012, \mnras, 421, 3543,
  \dodoi{10.1111/j.1365-2966.2012.20576.x}

\bibitem[{{Kotera} {et~al.}(2015){Kotera}, {Amato}, \&
  {Blasi}}]{2015JCAP...08..026K}
{Kotera}, K., {Amato}, E., \& {Blasi}, P. 2015, \jcap, 8, 026,
  \dodoi{10.1088/1475-7516/2015/08/026}

\bibitem[{{Lemoine} {et~al.}(2015){Lemoine}, {Kotera}, \&
  {P{\'e}tri}}]{2015JCAP...07..016L}
{Lemoine}, M., {Kotera}, K., \& {P{\'e}tri}, J. 2015, \jcap, 7, 016,
  \dodoi{10.1088/1475-7516/2015/07/016}

\bibitem[{{L{\'o}pez-Coto} {et~al.}(2018{\natexlab{a}}){L{\'o}pez-Coto},
  {Parsons}, {Hinton}, \& {Giacinti}}]{2018PhRvL.121y1106L}
{L{\'o}pez-Coto}, R., {Parsons}, R.~D., {Hinton}, J.~A., \& {Giacinti}, G.
  2018{\natexlab{a}}, Physical Review Letters, 121, 251106,
  \dodoi{10.1103/PhysRevLett.121.251106}

\bibitem[{{L{\'o}pez-Coto} {et~al.}(2018{\natexlab{b}}){L{\'o}pez-Coto},
  {Hahn}, {BenZvi}, {Dingus}, {Hinton}, {Nisa}, {Parsons}, {Greus}, {Zhang}, \&
  {Zhou}}]{2018APh...102....1L}
{L{\'o}pez-Coto}, R., {Hahn}, J., {BenZvi}, S., {et~al.} 2018{\natexlab{b}},
  Astroparticle Physics, 102, 1, \dodoi{10.1016/j.astropartphys.2018.04.003}

\bibitem[{{Malkov} {et~al.}(2013){Malkov}, {Diamond}, {Sagdeev}, {Aharonian},
  \& {Moskalenko}}]{2013ApJ...768...73M}
{Malkov}, M.~A., {Diamond}, P.~H., {Sagdeev}, R.~Z., {Aharonian}, F.~A., \&
  {Moskalenko}, I.~V. 2013, \apj, 768, 73, \dodoi{10.1088/0004-637X/768/1/73}

\bibitem[{{Malyshev} {et~al.}(2009){Malyshev}, {Cholis}, \&
  {Gelfand}}]{malyshev09}
{Malyshev}, D., {Cholis}, I., \& {Gelfand}, J. 2009, \prd, 80, 063005,
  \dodoi{10.1103/PhysRevD.80.063005}

\bibitem[{{Mignani} {et~al.}(2010){Mignani}, {Pavlov}, \&
  {Kargaltsev}}]{mignani10}
{Mignani}, R.~P., {Pavlov}, G.~G., \& {Kargaltsev}, O. 2010, \apj, 720, 1635,
  \dodoi{10.1088/0004-637X/720/2/1635}

\bibitem[{{Moderski} {et~al.}(2005){Moderski}, {Sikora}, {Coppi}, \&
  {Aharonian}}]{2005MNRAS.363..954M}
{Moderski}, R., {Sikora}, M., {Coppi}, P.~S., \& {Aharonian}, F. 2005, \mnras,
  363, 954, \dodoi{10.1111/j.1365-2966.2005.09494.x}

\bibitem[{{Moskalenko} \& {Strong}(1998)}]{1998ApJ...493..694M}
{Moskalenko}, I.~V., \& {Strong}, A.~W. 1998, \apj, 493, 694,
  \dodoi{10.1086/305152}

\bibitem[{{Olmi} \& {Bucciantini}(2019)}]{olmi19}
{Olmi}, B., \& {Bucciantini}, N. 2019, \mnras, 484, 5755,
  \dodoi{10.1093/mnras/stz382}

\bibitem[{{Posselt} {et~al.}(2015){Posselt}, {Spence}, \&
  {Pavlov}}]{2015ApJ...811...96P}
{Posselt}, B., {Spence}, G., \& {Pavlov}, G.~G. 2015, \apj, 811, 96,
  \dodoi{10.1088/0004-637X/811/2/96}

\bibitem[{{Posselt} {et~al.}(2017){Posselt}, {Pavlov}, {Slane}, {Romani},
  {Bucciantini}, {Bykov}, {Kargaltsev}, {Weisskopf}, \&
  {Ng}}]{2017ApJ...835...66P}
{Posselt}, B., {Pavlov}, G.~G., {Slane}, P.~O., {et~al.} 2017, \apj, 835, 66,
  \dodoi{10.3847/1538-4357/835/1/66}

\bibitem[{{Profumo}(2012)}]{profumo12}
{Profumo}, S. 2012, Central European Journal of Physics, 10, 1,
  \dodoi{10.2478/s11534-011-0099-z}

\bibitem[{{Profumo} {et~al.}(2018){Profumo}, {Reynoso-Cordova}, {Kaaz}, \&
  {Silverman}}]{2018PhRvD..97l3008P}
{Profumo}, S., {Reynoso-Cordova}, J., {Kaaz}, N., \& {Silverman}, M. 2018,
  \prd, 97, 123008, \dodoi{10.1103/PhysRevD.97.123008}

\bibitem[{{Rangelov} {et~al.}(2016){Rangelov}, {Pavlov}, {Kargaltsev},
  {Durant}, {Bykov}, \& {Krassilchtchikov}}]{rangelov16}
{Rangelov}, B., {Pavlov}, G.~G., {Kargaltsev}, O., {et~al.} 2016, \apj, 831,
  129, \dodoi{10.3847/0004-637X/831/2/129}

\bibitem[{{Reardon} {et~al.}(2016){Reardon}, {Hobbs}, {Coles}, {Levin},
  {Keith}, {Bailes}, {Bhat}, {Burke-Spolaor}, {Dai}, {Kerr}, {Lasky},
  {Manchester}, {Os{\l}owski}, {Ravi}, {Shannon}, {van Straten}, {Toomey},
  {Wang}, {Wen}, {You}, \& {Zhu}}]{reardon16}
{Reardon}, D.~J., {Hobbs}, G., {Coles}, W., {et~al.} 2016, \mnras, 455, 1751,
  \dodoi{10.1093/mnras/stv2395}

\bibitem[{{Recchia} {et~al.}(2018){Recchia}, {Gabici}, {Aharonian}, \&
  {Vink}}]{2018arXiv181107551R}
{Recchia}, S., {Gabici}, S., {Aharonian}, F.~A., \& {Vink}, J. 2018, arXiv
  e-prints.
\newblock \doarXiv{1811.07551}

\bibitem[{{Seta} {et~al.}(2018){Seta}, {Shukurov}, {Wood}, {Bushby}, \&
  {Snodin}}]{2018MNRAS.473.4544S}
{Seta}, A., {Shukurov}, A., {Wood}, T.~S., {Bushby}, P.~J., \& {Snodin}, A.~P.
  2018, \mnras, 473, 4544, \dodoi{10.1093/mnras/stx2606}

\bibitem[{{Silk} \& {Srednicki}(1984)}]{1984PhRvL..53..624S}
{Silk}, J., \& {Srednicki}, M. 1984, Physical Review Letters, 53, 624,
  \dodoi{10.1103/PhysRevLett.53.624}

\bibitem[{{Strong} {et~al.}(2007){Strong}, {Moskalenko}, \&
  {Ptuskin}}]{2007ARNPS..57..285S}
{Strong}, A.~W., {Moskalenko}, I.~V., \& {Ptuskin}, V.~S. 2007, Annual Review
  of Nuclear and Particle Science, 57, 285,
  \dodoi{10.1146/annurev.nucl.57.090506.123011}

\bibitem[{{Tang} \& {Piran}(2019)}]{2019MNRAS.484.3491T}
{Tang}, X., \& {Piran}, T. 2019, \mnras, 484, 3491,
  \dodoi{10.1093/mnras/stz268}

\bibitem[{{Venter} {et~al.}(2015){Venter}, {Kopp}, {Harding}, {Gonthier}, \&
  {B{\"u}sching}}]{2015ApJ...807..130V}
{Venter}, C., {Kopp}, A., {Harding}, A.~K., {Gonthier}, P.~L., \&
  {B{\"u}sching}, I. 2015, \apj, 807, 130, \dodoi{10.1088/0004-637X/807/2/130}

\bibitem[{{Vladimirov} {et~al.}(2012){Vladimirov}, {J{\'o}hannesson},
  {Moskalenko}, \& {Porter}}]{2012ApJ...752...68V}
{Vladimirov}, A.~E., {J{\'o}hannesson}, G., {Moskalenko}, I.~V., \& {Porter},
  T.~A. 2012, \apj, 752, 68, \dodoi{10.1088/0004-637X/752/1/68}

\bibitem[{{Y{\"u}ksel} {et~al.}(2009){Y{\"u}ksel}, {Kistler}, \&
  {Stanev}}]{yuksel09}
{Y{\"u}ksel}, H., {Kistler}, M.~D., \& {Stanev}, T. 2009, Physical Review
  Letters, 103, 051101, \dodoi{10.1103/PhysRevLett.103.051101}

\end{thebibliography}

\end{document}